\documentclass{ws-mpla}

\begin{document}

\markboth{H. Kamada, W. Gl\"ockle, H. Wita\l a, J. Golak, R.
Skibi\'nski, W. Polyzou, Ch. Elster}
{Lorentz boosted nucleon-nucleon T-matrix}

\catchline{}{}{}{}{}

\title{LORENTZ BOOSTED NUCLEON-NUCLEON T-MATRIX AND THE TRITON BINDING
ENERGY}

\author{\footnotesize H. KAMADA}

\address{Department of Physics, Faculty of Engineering, Kyushu Institute
of Technology, \\
1-1 Sensuicho Tobata, Kitakyushu 804-8550, Japan\\
kamada@mns.kyutech.ac.jp}

\author{W. GL\"OCKLE}

\address{
Institut f\"ur Theoretische Physik II, Ruhr-Universit\"at Bochum, \\
D-44780 Bochum, Germany}

\author{H. WITA\L A, J. GOLAK, R. SKIBI\'NSKI}
\address{M. Smoluchowski Institute of Physics, Jagiellonian University,
PL-30059 Krak\'ow, Poland}

\author{W. POLYZOU}
\address{Department of Physics and Astronomy, The University of Iowa,
Iowa City, IA 52242, USA}

\author{CH. ELSTER}
\address{Institute of Nuclear and Particle Physics, and Department of
Physics, Ohio University, Athens, Ohio 45701, USA}

\maketitle

\pub{Received (Day Month Year)}{Revised (Day Month Year)}

\begin{abstract}
The phase equivalent relativistic NN potential, 
which is related by a
nonlinear equation to the original
nonrelativistic  potential,
is used 
to construct the mass operator (rest Hamiltonian)
of the 
3-nucleon system.
Employing the CD Bonn NN  potential,
the solution of the relativistic 3N
Faddeev equation for $^3$H shows slightly
less binding energy than the corresponding nonrelativistic result.
The effect of the Wigner spin rotation on the binding is very small.
\end{abstract}

\ccode{PACS Nos.:21.45.+v, 24.70.+s, 25.10.+s, 25.40.Lw }

\section{Introduction}

Considerable experimental effort has been made in measuring
proton-deuteron (pd) scattering
\cite{Abfal,Sekiguchi,Hatanaka,Ermisch05,Kistryn05}
cross sections
at intermediate
energies.
For up to 300~MeV proton energy those data have been analyzed 
with rigorous three-nucleon (3N) Faddeev calculations~\cite{witala}
based on the CD-Bonn potential~\cite{Machleidt} and the
Tucson-Melbourne 3N force (3NF) \cite{Coon}.  Theoretical predictions
based on 2N forces alone are not sufficient to describe the data above
about 100 MeV.  Some of those defects are known as 
the 
Sagara discrepancy~\cite{Sagara,Nemoto,witala2}.  Though 3NF effects are
already seen below 100 MeV, they increase significantly above that
energy.  However, presently available 3NF's only partially improve the
description of cross section data and spin observables.  Since most of
the cited calculations are based on the non-relativistic formulation
of the Faddeev equations~\cite{PR}, one needs to question if in the
intermediate energy regime a Poincar{\'e} invariant formulation is
more adequate.

There are two basic approaches to a relativistic formulation of the 3N
problem. One is a manifestly covariant scheme linked to a field
theoretical approach~\cite{Stadler}, the other is based on
an exact realization of the symmetry of the Poincar\'e group
in three nucleon quantum mechanics~\cite{Coester}. 
We employ the second approach, where the 
mass operator (rest energy operator) 
consists of relativistic kinetic energies together with two-
and many-body interactions including their boost
corrections~\cite{Bakamjian}.

The first attempt in solving the relativistic Faddeev equation for the
3N bound state based on second approach has been carried out
in~\cite{Gloeckle}, resulting in a decrease of the binding energy
compared to the nonrelativistic result.  On the other hand, similar
calculations based on the field theory approach~\cite{Stadler}
increase it. These contradictory results require more
investigation.
In the following we
summarize the results of
our calculations based on the second approach:
in Section 2 we introduce the relativistic 2N potential, 
in Section 3 we present the 2N t-matrix, which fulfills the relativistic
boosted Lippmann-Schwinger (LS) equation,
and in Section 4 we give numerical 
results for the triton binding energy based on the Poincar{\'e} invariant
Faddeev equation.

\section{The Relativistic Potential}

Modern meson theoretical NN potentials, e.g. charge dependent
Bonn Potential (CD-Bonn)~\cite{Machleidt},
are derived from a  relativistic Lagrangian, then cast into a three-dimensional
form using the Blankenbeclar-Sugar equation, which by
kinematical redefinitions can be written in the form of a standard
nonrelativistic LS equation, which in partial wave decomposed form reads
\begin{eqnarray}
t(p,p'; \frac{ p'^ 2}{m}) = v(p,p') + \int_0^\infty { v(p,p')
t(p'',p';\frac{ p'^ 2}{m})
 \over \frac{ p'^ 2}{m}- \frac{p''^2}{m} +i\epsilon}   {p''}^2 dp''.
\label{eq1}
\end{eqnarray}
The corresponding relativistic LS  equation is given as
\begin{eqnarray}t^{r}(p,p';E) = v^{r}(p,p') + \int_0^\infty {
v^{r}(p,p'') t^{r}(p'',p'; E)
\over E - 2\sqrt{ {p''}^2+m^2 } +i\epsilon}   {p''}^2 dp''
\label{eq5}
\end{eqnarray}
where
\begin{eqnarray}
E=2\sqrt{p'^2+m^2}
\end{eqnarray}

In  the relativistic Faddeev equation one needs $t^r$
off-the-energy-shell.
According to \cite{Coester2} there is a direct operator relation between 
the nonrelativistic $v$ and the relativistic $v^r$:
\begin{equation}
4 m~ {\hat v}=  2\sqrt{ {\hat p}^2+ m^2} \; {\hat v}^r +2 {\hat v}^r
\sqrt{ {\hat p}^2+m^2} + ({\hat v}^r)^2
\label{eq6}
\end{equation}
In a momentum representation this leads to  
\begin{eqnarray}
\lefteqn{4 m \; v(p,p') =} \cr 
& & v^r (p,p')  \left( 2\sqrt{p^2+m^2} + 
2\sqrt{ p'^2 +m^2}  \right) +
\int_0^\infty  dp'' \; p''^2 \; v^r(p,p'') \; v^r(p'',p') .
\label{eq7}
\end{eqnarray}
This is the nonlinear relation between the relativistic potential $v^r$
and the nonrelativistic potential  $v$ from Eq.~(\ref{eq1}),
which has recently been solved~\cite{Kamada}. The resulting
on-shell-t-matrix $t^r$ is
on-shell identical to the t-matrix $t$ from Eq.~(\ref{eq1}).

\section{The Lorentz Boosted T-matrix}
Cluster properties require that the energy is additive.  Because of
the non-linear relations between the mass and energy in special
relativity, the additivity of energies in the rest frame implies a
non-linear relation between the two-body interactions in the two and
three-body mass operators ~\cite{Coester}.  We call the two-body interaction in 
the three-body mass operator the ``boosted potential''.
\begin{eqnarray}
\hat {v_q^r}\equiv
 \sqrt{ \left[ 2 \sqrt{ {\hat p}^2 + m^2}  + \hat {v^r } \right]^2  + q^2 }
-\sqrt{\left[ 2 \sqrt{{\hat p}^2 + m^2}\right]^2  + q^2 },
\label{eq8}
\end{eqnarray}
where the spectator momentum $q$ in the 3-body center of mass is simultaneously the
negative total momentum of the pair.
Using  Eq.~(\ref{eq6}) this can be rewritten as~\cite{Kamada}
\begin{eqnarray}
4 m \; v(p,p')&=& v^r_q (p,p') \left( \sqrt{4(p^2+m^2)+q^2} + \sqrt{4({p'}^2
+m^2) +q^2} \right) \cr 
&+& \int _0 ^\infty dp '' \; {p''}^2 \;
v^r_q(p,p'') v^r_q(p'',p') .
\label{eq9}
\end{eqnarray}
Thus one can obtain $v^r_q$ by the same technique~\cite{Kamada} as $v^ r$. 
The boosted off-shell t-matrix is the solution of the LS
equation
\begin{eqnarray}
t^{r}_q(p,p';E_q) &=& v^{r}_q(p,p') \cr
 &+& \int_0^\infty { v^{r}_q(p,p')~
 t^{r}_q(p'',p'; E_q) \over \sqrt{ 4({k}^2+m^2) +q^2 } - \sqrt{
4({p''}^2+m^2)+q^2 } +i\epsilon}   {p''}^2 dp''  .
\label{eq10}
\end{eqnarray}
with 
$E_q=\sqrt{ 4({k}^2+m^2) +q^2 }$.

\begin{figure}[ph]
\centerline{\psfig{file=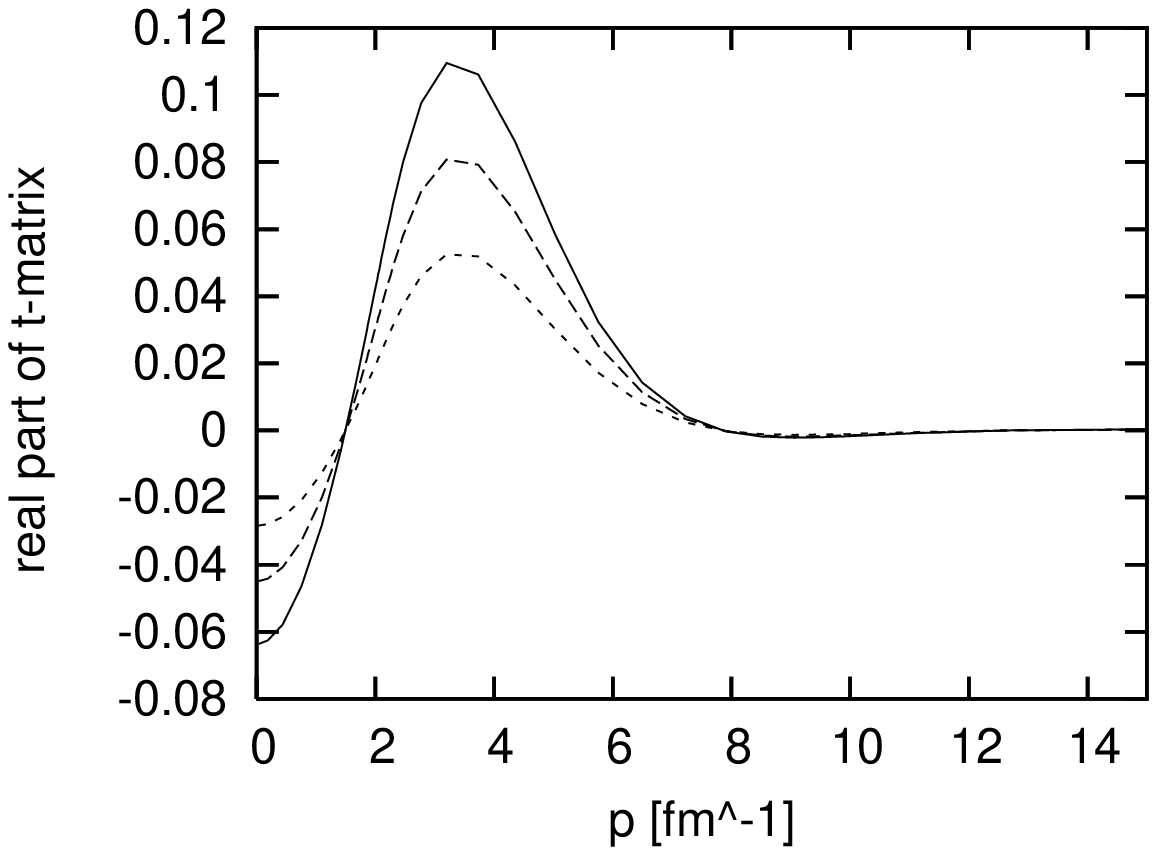,width=1.9in,angle=-90}
\psfig{file=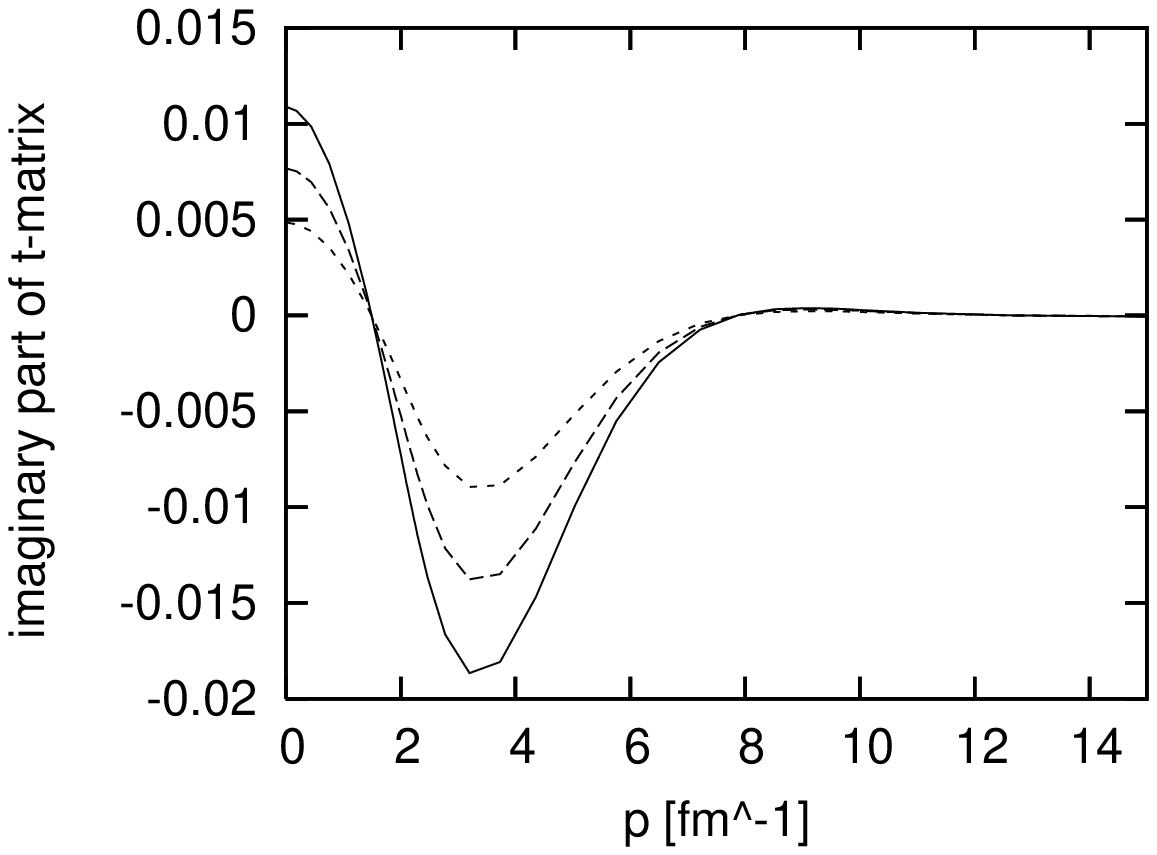,width=1.9in,angle=-90}
}
\vspace*{8pt}
\caption{
The boosted half-on-the-mass-shell t-matrix of the CD-Bonn potential at
$E_{lab}$=350 MeV. The left and right plots are real  and imaginary
parts, respectively. The solid, dashed and dotted lines are related to
the boosting momentum $q$= 0, 10 and 20 $\rm fm^{-1}$,
respectively. \protect\label{figtmat}}
\end{figure}

In Fig. \ref{figtmat} we display the boosted half-shell
($p'=k$) t-matrix of the CD-Bonn~\cite{Machleidt} potential at
$E_{lab}$=350 MeV for three different spectator momenta $q$. 
The magnitude of the t-matrix gradually decreases with increasing the
boost momentum $q$. It can be shown~\cite{Keister:2005eq,Lin:2008sy}
that 
the half-shell t-matrices  $t^r_q (p,p'=k;E_k)$ and $t^r (p,p'=k;E_k)$ are 
related by simple factors
\begin{eqnarray}
t^r_q(p,k; E_k)={2\sqrt{p^2+m^2}+2\sqrt{k^2+m^2} \over \sqrt{ 4({p}^2+m^2)
+q^2 } + \sqrt{ 4({k}^2+m^2)+q^2 }
} t^r (p,k; E_k).
\label{eq11}
\end{eqnarray}
Solving Eqs.~(\ref{eq5}) and (\ref{eq10}) to obtain $t^r$ and $t^r_q$
independently we numerically confirmed  the relation (\ref{eq11}) with
high precision.
It is the factor on front of $t^r$ in the right hand side of Eq.~(\ref{eq11}) 
which  attenuates the amplitude of
the t-matrix with increasing  $q$. It can also be shown~\cite{Keister:2005eq} that
the relativistic half-shell t-matrix $t^r$ is related to the corresponding 
nonrelativistic one via
\begin{equation}
t^r(p,k; E=2\sqrt{k^2+m^2}) = \frac{4m}{2\sqrt{k^2+m^2} +2\sqrt{p^2+m^2}} 
                 t(p,k;k^2/m).
\end{equation}

The explicit  
construction of first $v^r$ and then $t^r$ is equivalent to obtaining $t^r$ via
resolvent equations as suggested in~\cite{Keister:2005eq} and carried out 
in~\cite{Lin:2007kg,Lin:2008sy}.

\section{The Triton Binding Energy}
The relativistic bound state Faddeev
equation was solved
using the boosted t-matrix $t_q^r$. 
In Table \ref{table1} the results for the triton binding energy using the CD-Bonn
potential as input  are shown. The  triton binding
energy obtained from the relativistic calculation
is about 100~keV smaller compared to the one calculated nonrelativistically. 
This value
is significantly smaller than a previously published 
result~\cite{Kamada2} in which a reduction of the triton binding energy
by about 400~keV was given.  
The reason for this overestimation of a relativistic effect on the
binding energy can be attributed to a different construction of the 
relativistic off-shell t-matrix $t^r$. The scaling transformation
employed in~\cite{Kamada2} does not keep the 2N scattering data invariant
as function of the 2N c.m. momentum.

\begin{table}[h]
\tbl{
The theoretical predictions of the trition binding energies resulting 
from the solutions of the nonrelativistic (first row) and
relativistic (second row) Faddeev equations as function of the number
 of partial waves taken into account. 
The last line indicates the absolute difference between 
the nonrelativistic and relativistic result. In the calculations
only the np force of the CD-Bonn potential was used. Unit is in MeV.
}
{\begin{tabular}{@{}ccccc@{}} \toprule
    & 5ch  (S-wave)    & 18ch ($j_{max}=2$) &  26ch ($j_{max}=3$) &
34ch   ($j_{max}=4$)  \\
\colrule
 nonrel. &   -8.331  & -8.220 &  -8.241 & -8.247   \\
rel.     &   -8.219  & -8.123 &  -8.143 & -8.147   \\
diff.    &   ~0.112  & ~0.107 &  ~0.098 & ~0.100   \\
\botrule
\end{tabular}}
\label{table1}
\end{table}

We also included the Wigner spin rotation as outlined in
\cite{Witala08}. Thereby the  the Balian-Brezin method\cite{Balian} in
handling the permutations is quite useful. In Table \ref{table2}  the
triton binding energies are shown allowing
charge independence breaking (CIB)\cite{Witala91} \cite{} and Wigner
spin rotations. 
Wigner spin rotation effects reduce the binding energy
by only about 2 keV.

\begin{table}[h]
\tbl{The theoretical predictions for the relativistic and
nonrelativistic triton binding energies in MeV. All numbers are 34
channels results. The second column is the same as the last column in
Table \ref{table1}. The results in the  third column take charge
dependence\protect\cite{Witala91} into account. In addition the result
of the fourth column contains also Wigner spin rotation effects. }
{\begin{tabular}{@{}ccccc@{}} \toprule
    & ~~~np force only~~~    & ~~~np+nn forces~~~ &  ~~~Wigner
rotation~~~  &  diff.  \\
\colrule
 nonrel. &   -8.247  & -8.005 &  - & -  \\
rel.     &   -8.147  & -7.916 &  -7.914 & -0.002   \\
diff.    &   ~0.100  & ~0.089 &   -  &  -   \\
\botrule
\end{tabular}}
\label{table2}
\end{table}

\section{Summary and Outlook}

A phase-shift  equivalent 2N potential $ \hat v^ r$ in the relativistic
2N 
Schr\"odinger 
equation is related to the potential $v$ in the
nonrelativistic 
Schr\"odinger 
equation by the nonlinear relation given
in Eq.~(\ref{eq6}). The boosted potential $\hat v^ r_q$ is
related to $ \hat v^ r$ by a similar expression, Eq.~(\ref{eq8}). 
With these potentials we generate
the relativistic fully-off-shell t-matrix $ t^ r_q$,
which enters into the relativistic Faddeev equation. 
We solve the relativistic bound state Faddeev equation and compare
the binding energy for the triton with the one obtained from a 
nonrelativistic calculation
with the same input interaction.
We find that the difference between 
the two calculations is only about 90~keV including CIB, where the relativistic
calculation gives slightly less binding. Taking Wigner spin rotations into
account in the relativistic calculation reduces the binding energy by a very small amount, 
$\approx$~2~keV, indicating that Wigner rotations of the spin have essentially no effect
on the predicted value of the binding energy.

Applications to the 3-body continuum are in progress.
Recently~\cite{Witala08} the formulation lined out above has been used
to study the low energy $A_y$ puzzle in neutron-deuteron scattering.
Details are presented by Wita\l a in this conference. In the
intermediate energy regime the formulation has been applied to
exclusive proton-deuteron scattering cross sections at
508~MeV~\cite{Lin:2007kg,Lin:2008sy} based on a formulation of of the
Faddeev equations which does not employ a partial wave decomposition.
The approach can also be extended and applied to electromagnetic
processes\cite{Golak,Golak07}.

\section*{Acknowledgments}
This work was partially supported by the 2008-2011 polish science funds as a
research project No. N N202 077435. 
It was also partially supported by
the Helmholtz Association
through funds provided to the virtual institute ``Spin and strong
QCD''(VH-VI-231). The numerical calculations were performed on the IBM
Regatta p690+ of
the NIC in J\"ulich, Germany.

\end{document}